\documentclass[prl,superscriptaddress,aps, twocolumn, reprint, showpacs, longbibliography, footnoography]{revtex4-2}
\usepackage{graphicx}
\usepackage{amssymb}   % for math
\usepackage{amsfonts}
\usepackage{amsmath}
\usepackage{dsfont}
\usepackage{natbib}
\usepackage{dcolumn}   % needed for some tables
\usepackage{bm}        % for math
\usepackage[mathscr]{eucal}
\usepackage[dvipsnames]{xcolor}
\usepackage[colorlinks, linkcolor=OrangeRed,citecolor=RoyalBlue,urlcolor=NavyBlue]{hyperref}
\usepackage[all]{hypcap} % let hyperlinks correctly point to figures rather than their captions;
\usepackage{mathtools}
\usepackage{dsfont}

\definecolor{myblue}{rgb}{0,0,0.75}

\newcommand{\ep}{\varepsilon}

\newcommand\mean[1]{\ensuremath{\left\langle#1\right\rangle}}

\newcommand\lrp[1]{\left(#1\right)}
\newcommand\lrb[1]{\left[#1\right]}

\newcommand{\be}{\begin{equation}}
\newcommand{\ee}{\end{equation}}
\def\ba{\begin{aligned}}
\def\ea{\end{aligned}}
\newcommand{\bea}{\begin{eqnarray}}
\newcommand{\eea}{\end{eqnarray}}
\def\bes{\begin{subequations}}
\def\ees{\end{subequations}}
\def\bal{\begin{align}}
\def\eal{\end{align}}

\renewcommand{\vec}[1]{{\bf #1}}

\usepackage{soul}
\usepackage{ulem} % needed for \sout-command

\usepackage{graphicx}
\usepackage{amsmath,amssymb,bm}
\usepackage{enumerate}
\usepackage{braket}        % Dirac notation

\begin{document}
\title{Ergodicity-breaking phase diagram and fractal dimensions in long-range models with generically correlated disorder}

\author{Shilpi Roy}
\affiliation{Department of Physics, Indian Institute of Technology Guwahati, Guwahati-781039, Assam, India}

\author{Saurabh Basu}
\affiliation{Department of Physics, Indian Institute of Technology Guwahati, Guwahati-781039, Assam, India}

\author{Ivan~M.~Khaymovich}
\email{ivan.khaymovich@gmail.com}
\affiliation{Nordita, Stockholm University and KTH Royal Institute of Technology Hannes Alfv\'ens v\"ag 12, SE-106 91 Stockholm, Sweden}
\affiliation{Institute for Physics of Microstructures, Russian Academy of Sciences, 603950 Nizhny Novgorod, GSP-105, Russia}

%\date{\today}%
\begin{abstract}
Models with correlated disorder are rather common in physics. In some of them, like the Aubry-Andr\'e (AA) model, the localization phase diagram can be found from the (self)duality with respect to the Fourier transform. In others, like the all-to-all translation-invariant Rosenzweig-Porter (TI RP)  ensemble or the Hilbert-space structure of the many-body localization, one needs to develop more sophisticated and usually phenomenological methods to obtain the localization transition.
In addition, such models contain not only localization but also the ergodicity-breaking transition, giving way to the non-ergodic extended phase of states with non-trivial fractal dimensions $D_q$.
In this work, we suggest a method to calculate both the above transitions and a lower bound to the fractal dimensions $D_2$ and $D_\infty$, relevant for the physical observables.
In order to verify this method, we apply it to the class of long-range (self-)dual models, interpolating between AA and TI RP ones via both power-law dependencies of the on-site disorder correlations and the hopping terms, and, thus, being out of the validity range of the previously developed methods.
We show that the interplay of the correlated disorder and the power-law decaying hopping terms leads to the emergence of the two types of fractal phases in the entire range of parameters, even without having any quasiperiodicity of the AA potential.
The analytical results of the above method are in full agreement with the extensive numerical calculations.
\end{abstract}

\maketitle

%%%%%%%%%%%%%%%%%%%%%%%%%%%%%%%
\textit{Introduction --}
%%%%%%%%%%%%%%%%%%%%%%%%%%%%%%%
The concept of quantum ergodicity and its breaking is not only of a fundamental interest to understand the many-body localization phenomenon~\cite{Basko06,gornyi2005interacting,Pal2010,Alet_CRP,Abanin_RMP}, but also has the direct applications to the superconductivity enhancement (see, e.g.,~\cite{Feigelman2007SC-enhance,Feigelman2010AoP,Petrovic2016disorder} and many others). Additionally, it has implications in quantum-algorithm speed-ups~\cite{smelyanskiy2018non,kechedzhi2018efficient},
and the black-hole description~\cite{micklitz2019non,Kamenev-talk-1,Kamenev-talk-2}. Being delocalized but occupying measure zero of all available sites, the (multi)fractal states~\cite{EversMirlin2008} constitute the non-ergodic extended phase of matter. The latter generalizes the concept of Anderson localization~\cite{Anderson1958} by including the ergodic-to-nonergodic transition~\cite{Kravtsov_NJP2015} in addition to the localization one.

The realization of a multifractal phase typically requires all-to-all coupling, such as in the Rosenzweig-Porter (RP) model~\cite{RP,Kravtsov_NJP2015,Biroli_RP,Ossipov_EPL2016_H+V,vonSoosten2017non,Monthus2017multifractality,BogomolnyRP2018,LN-RP_RRG,LN-RP_WE,Biroli2021levy,LN-RP_K(w),Buijsman2022circular,Venturelli2023replica,DeTomasi2022nH-RP,sarkar2023fract-RP}. Alternatively, quasiperiodicity of the on-site disorder potential~\cite{AA}, full~\cite{Nosov2019correlation,Nosov2019mixtures,Deng2016Levy,Deng2018Duality,Deng2022AnisBM} or partial~\cite{Kutlin2021emergent,Motamarri2022RDM,Tang2022nonergodic} correlations in the kinetic term are needed for realizing a multifractal phase (with only a few exceptions like~\cite{Das2022beta-ens,Das2023beta-ens_IMK}).
While the description of non-ergodic extended phases in long-range models, even with partial correlations~\cite{Kutlin2021emergent,Motamarri2022RDM,Tang2022nonergodic}, has been well-developed with mathematical physics rigor~\cite{vonSoosten2017non,vonSoosten2017phase,Venturelli2023replica}, disordered models with partially correlated on-site potentials are more subtle.

The correlated on-site disorder in quantum systems is widely used to realize the localization-delocalization transition in low-dimensional systems~\cite{AA}, inducing non-ergodic extended phases of matter in various systems. These phases manifest themselves in tight-binding~\cite{Duthie_Roy2022_MF-correlated,Goncalves-Ribeiro2023incommensurability,Goncalves-Ribeiro2023critical,gao2023experimental}, long-range~\cite{Liu2015AA=2dHarper,Gopalakrishnan2017,Deng2019Quasicrystals}, or flat-band models~\cite{Chalker2010flat-band,Danieli2015_flat-band,Ahmed2022flat-band,Lee-Flach2023flat-band_MF}, as well as in superconducting~\cite{Cai2013AA+p-wave,DeGottardi2013AA+p-wave,Wang2016AA+p-wave,Fraxanet2021AA+p-wave,Fraxanet2022AA+p-wave} and Floquet-driven systems~\cite{Roy2018,Sarkar2021_Floquet_MF,Ray_Floquet_MF,Sarkar2021signatures,goncalves2023quasiperiodicity}. Many of these models are based on a quasiperiodic fully-correlated on-site potential proposed by Aubry and André in~\cite{AA}, which exhibits a hidden fractal structure~\cite{Kravtsov2023Random_cantor,Goncalves-Ribeiro2023critical,Goncalves-Ribeiro2023incommensurability,sarkar2023fract-RP,goncalves2023quasiperiodicity}. This model makes use of the duality between real and momentum spaces via the Fourier transform~\cite{AA,Goncalves-Ribeiro2023incommensurability,Goncalves-Ribeiro2023critical}, which has also been generalized to long-range models with uncorrelated disorder~\cite{Nosov2019correlation}. Description of other systems utilizes their mapping to the $2$d Harper model~\cite{Liu2015AA=2dHarper,Danieli2015_flat-band,Ahmed2022flat-band,Lee-Flach2023flat-band_MF}.

In this Letter, we present another approach that goes beyond the above-mentioned techniques, enabling us to describe both the Anderson and ergodic transitions in models featuring partially correlated on-site disorder potential without relying on quasiperiodicity. We develop a technique to determine the position of the localization transition, which generalizes the concept of resonance-counting spatial renormalization group~\cite{Levitov1989,Levitov1990,Mirlin2000RG_PLRBM,Burin1989,Kutlin2020_PLE-RG} to account for partially correlated on-site disorder. Furthermore, we calculate the lower bound for the fractal dimensions necessary to identify the location of the ergodic transition.

\begin{figure}[h!]
\label{Fig1}
    \includegraphics[width=1\columnwidth]{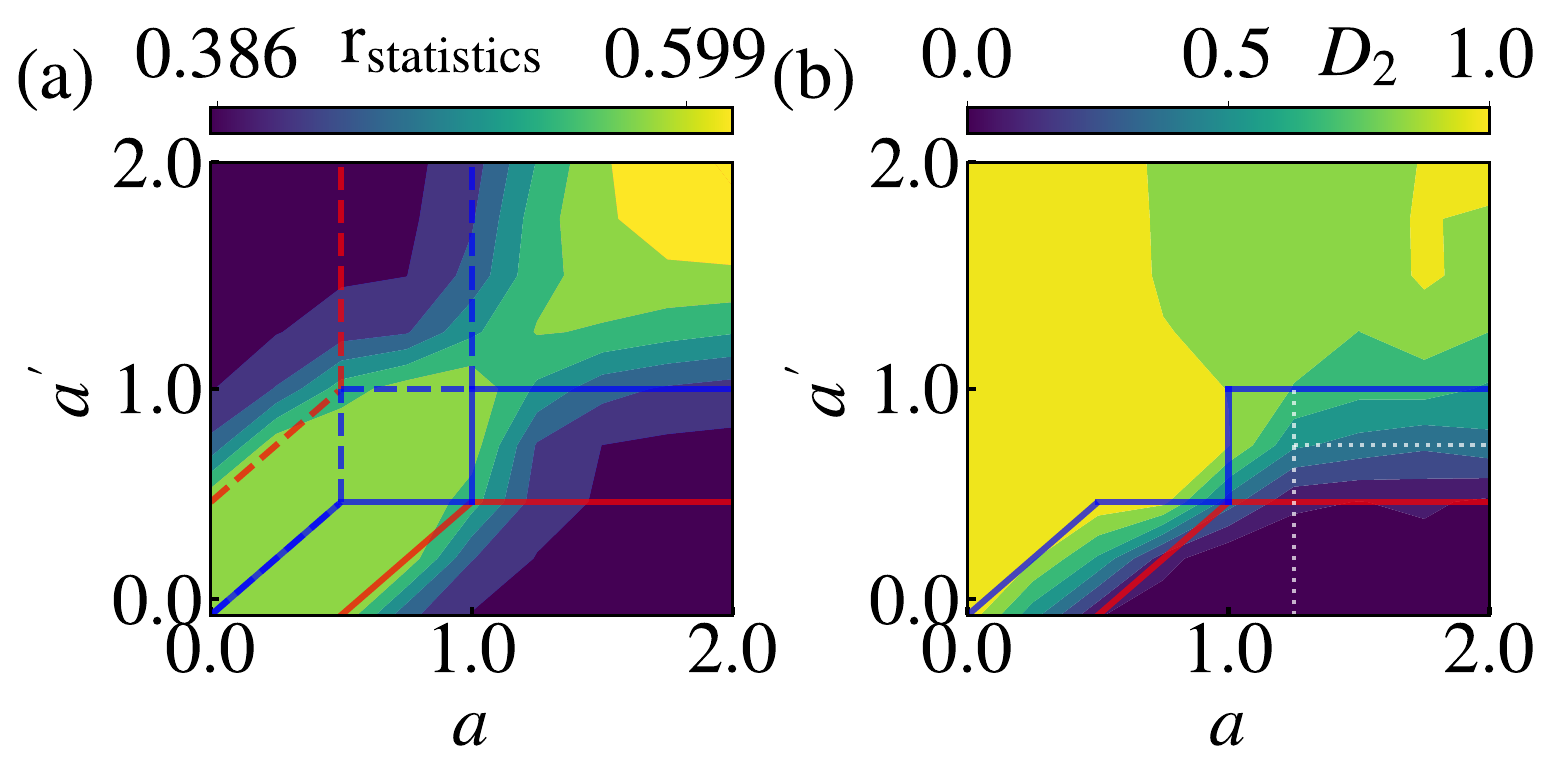}
    \caption{\textbf{Phase diagram of the self-dual model with power-law decaying hopping in real and momentum spaces}: (a)~mean spectrum gap ratio $r$, Eq.~\eqref{eq:r-stat}, and (b)~eigenstate real-space fractal dimension $D_2$, Eq.~\eqref{eq:IPR_q}, both averaged over the spectral bulk, vs the hopping decay rates in real $a$ and momentum $a'$ spaces. Blue (red) solid [dashed] lines show the ergodic (localization) transitions in real [momentum] space. All the data in this and both other figures are calculated for $2^9\leq N\leq 2^{14}$ with $500$ disorder realizations and extrapolated for the fractal dimensions. Note that the phases, ergodic in both spaces, $a,a^\prime>1$ and $0.5<a,a^\prime<1$, have strong finite-size effects both in $D_2$ and $r$.
    }
\end{figure}

We verify our method on an exemplary model that interpolates between the self-dual Aubry-Andr\'e (AA) model and its long-range cousin, the translation-invariant (TI) RP ensemble~\cite{Nosov2019correlation}, with uncorrelated disorder and all-to-all TI coupling. Our model contains power-law correlated on-site disorder and TI hopping, which decays with distance from the diagonal following a power-law. By studying this model, we are able to explore the phase diagram, as shown in Fig.~\ref{Fig1}.

%%%%%%%%%%%%%%%%%%%%%%%%%%%%%%%%%%%%%%%%%%%%%%%%%%%%%%%%%%%%%%
\textit{Resonance counting in long-range models -- }
%%%%%%%%%%%%%%%%%%%%%%%%%%%%%%%%%%%%%%%%%%%%%%%%%%%%%%%%%%%%%%
Among the models that exhibit Anderson transition, there are two paradigmatic ones, the $3$d Anderson model (3dAM)~\cite{Anderson1958} and the power-law random banded matrix (PLRBM) one~\cite{Levitov1989,Levitov1990,PLRBM}. These models are described by the Hamiltonian in the matrix notation of the $d$-dimensional real space of $N$ sites ($\vec{n}$, $\vec{m}$),
\be\label{eq:ham}
H_{\vec{mn}} = \ep_{\vec{n}} \delta_{\vec{mn}} + J_{\vec{mn}} \ ,
\ee
characterized by the i.i.d. random on-site disorder $\ep_{\vec{n}}$, $\mean{\ep_{\vec{n}}}=0$, $2\mean{\ep_{\vec{n}}\ep_{\vec{m}}} = \delta_{\vec{mn}} W^2$, and the random independent hopping $J_{\vec{mn}}$, $\mean{J_{\vec{mn}}}=0$, power-law decaying on average $\mean{J_{\vec{mn}}J_{\vec{kl}}^*} = \delta_{\vec{mk}}\delta_{\vec{nl}}/|\vec{m-n}|^a$.
Here $\delta_{\vec{mn}}$ is a Kronecker delta.
Unlike this, the model in the focus of our work, depicted in the phase diagram in Fig.~\ref{Fig1}, has correlated on-site disorder [Eqs.~\eqref{eq:eps_n},~\eqref{eq:J_p}]. Before delving into this model, let us briefly revisit the case of uncorrelated disorder.

Both models, 3dAM ($a\to\infty$, $d=3$) and PLRBM ones ($W=1$, any $a$, $d$), show the localization transition vs disorder ($W$ for 3dAM, $a$ for PLRBM), which has been described using the so-called resonance counting~\cite{Anderson1958,Levitov1989} and the spatial renormalization group~\cite{Levitov1990,Mirlin2000RG_PLRBM}.
Let's focus on the PLRBM case, where this approach straightforwardly shows the transition at $a=d$.
Indeed, the standard resonance counting is given by the following algorithm:

In the spatial perturbation theory and the renormalization-group analysis, terms like ${J_{\vec{mn}}}/\lrp{\epsilon_{\vec{n}}-\epsilon_{\vec{m}}}$ arise and are assumed to be small for perturbation-theory convergence. This assumption holds when the following condition is satisfied:
\begin{gather}
\left|\frac{J_{\vec{mn}}}{\epsilon_{\vec{n}}-\epsilon_{\vec{m}}}\right|<1 \ .
\end{gather}
Any violation of this condition, called a resonance, should be treated via a degenerate perturbation theory.

A certain state with the wave-function maximum at site $\vec{m}$ is Anderson-localized if the number of resonances is finite. Otherwise, the above approach breaks down, usually leading to delocalization.
In order to calculate the number of resonances, one should consider all possible sites at a distance $\sim R$ from the initial site $\vec{m}$. There are $R^d$ such sites in $d$-dimensional lattice, with i.i.d. random $\epsilon_{\vec{m+R}}$, distributed in the window of width $\sim W$. Therefore, the corresponding level spacing (between adjacent levels) is given by
\begin{gather}\label{eq:delta_R}
\delta_R = \frac{\sqrt{\langle(\epsilon_{\vec{m+R}}-\epsilon_{\vec{m}})^2\rangle}}{R^d}.
\end{gather}
Comparing $\delta_R$ with the corresponding hopping term $J_{m,m+R} \sim 1/R^a$, one can find the transition at $a=d$. Indeed, at all $a>d$ beyond a certain distance $R > R_* = W^{-1/(a-d)}\sim O(1)$ at which the last resonance appear
\be\label{eq:R*}
\delta_{R_*}=J_{R_*}
\ee
one has $\delta_R > J_R$, i.e., typically no resonances and localization~\footnote{Here, one should mention that in some models with correlated $J_{mn}$ like~\cite{Burin1989,Malyshev2000,Deng2018Duality,Nosov2019correlation} there might be a transition over $W$ for $a>d$.}.
On the other hand, for $a<d$ beyond a finite distance, $R > R^{**}=W^{1/(d-a)} \sim O(1)$, one has $\delta_R < J_R$, i.e., the number of resonances, given by the ratio $J_R/\delta_R\sim R^{d-a}$, grows with $R$. Therefore the states are delocalized.

The spatial renormalization group~\cite{Levitov1990,Mirlin2000RG_PLRBM}, in addition to the above resonance counting, takes into account the effects of hybridization of resonance site pairs via the degenerate perturbation theory.

%%%%%%%%%%%%%%%%%%%%%%%%%%%%%%%%%%%%%%%%%%%%%%%%%%%%%%%%%%%%%%
\textit{Resonance counting with correlated disorder -- }
%%%%%%%%%%%%%%%%%%%%%%%%%%%%%%%%%%%%%%%%%%%%%%%%%%%%%%%%%%%%%%
As a main result of this paper, we generalize
the above-mentioned resonance counting approach to the correlated diagonal disorder $\ep_{\vec{n}}$.
First of all, for the correlators $\mean{\ep_{\vec{m}}\ep_{\vec{n}}}$, which depend only on the vector $\vec{m-n}$, as shown in %~Sec.~I of~\cite{SM},
Appendix~\ref{App_Sec:general_correlations},
is given in general by a Fourier transform of certain independent random  $\bar{J}_{\vec{p}}$
\begin{equation}\label{eq:eps_n}
\epsilon_{\vec{n}}=\sum\nolimits_{\vec{p}} \bar{J}_{\vec{p}} e^{{i\frac{2\pi}{N}}\vec{p n}} \ , \quad \mean{\bar{J}_{\vec{p}}\bar{J}_{\vec{q}}^*} = \delta_{\vec{pq}} F(\vec{p}) \ .
\end{equation}
In our exemplary model, we will focus on %the power-law decaying $\bar{J}_p$
\be\label{eq:J_p} % CAN BE SHORTENED BY PUTTING INLINE!!
\bar{J}_p  = {\bar{j}_{p}}/{|p|^{a^{\prime}}} \text{, with } \mean{\bar{j}_{p}}=0,\; \mean{\bar{j}_{p}^2}=1  \ .
\ee
In a general case, if the series~\eqref{eq:eps_n} does not absolutely converge, i.e., is determined by the extensive number of $\bar{J}_{\vec{p}}$, that is not fat-tail distributed, the variables $\epsilon_{\vec{n}}$ have Gaussian distribution. Thus, their differences $\epsilon_{\vec{n+R}}-\epsilon_{\vec{n}}$ are also Gaussian-distributed with a certain variance $W_R^2 = \mean{\lrp{\epsilon_{\vec{n+R}}-\epsilon_{\vec{n}}}^2}$, straightforwardly calculated from Eq.~\eqref{eq:eps_n}.

The latter allows one to apply the resonance counting method, described above with the $R$-dependent width $W_R$ in Eq.~\eqref{eq:delta_R}.
Thus, the convergence of the number of resonances with the system size $N$ will determine the localized phase of the model, while its first divergence gives the Anderson transition location in the parameter space.

%%%%%%%%%%%%%%%%%%%%%%%%%%%%%%%%%%%%%%%%%%%%%%%%%%%%%%%%%%%%%%
\textit{General fractal dimension estimates -- }
%%%%%%%%%%%%%%%%%%%%%%%%%%%%%%%%%%%%%%%%%%%%%%%%%%%%%%%%%%%%%%
Unlike PLRBM model, with only localized (ergodic extended) states for $a>d$ ($a<d$), in the models with correlated disorder, Eq.~\eqref{eq:eps_n}, the non-ergodic extended phases with fractal eigenstates $\psi_{E_l}(\vec{n})$ may appear.
%\rev{In the models with correlated disorder [Eq.~\eqref{eq:eps_n}], non-ergodic extended phases with fractal eigenstates $\psi_{E_l}(\vec{n})$ may appear, unlike the PLRBM model where only localized (ergodic extended) states appear for $a>d$ ($a<d$).}
In order to characterize such wave-function structures, one needs to calculate so-called fractal dimensions $D_q$ from the $q$-order inverse participation ratio (IPR) as follows~\cite{EversMirlin2008}:
\begin{gather}\label{eq:IPR_q}
I_q = \sum_{\vec{n}} |\psi_{E_l}(\vec{n})|^{2q} \sim N^{(1-q) D_q} \ .
\end{gather}
The fractal dimensions are limited from above and below and form a monotonic in $q$ function
$1\geq D_q \geq D_{q+1} \geq 0$.

The minimal fractal dimension $D_{q\to \infty}$ characterizes the scaling of the wave-function  maximum $\max_n |\psi_{E_l}(\vec{n})|^2 \sim N^{-D_\infty}$. Due to the normalization condition, Eq.~\eqref{eq:IPR_q} at $q=1$, it can be estimated as the number of sites where the wave function coefficients are of the same order as at the maximum:
\begin{gather}\label{eq:D_infty}
N^{D_\infty}\sim \#\{n: |\psi_{E_l}(\vec{n})|^2/\max_{\vec{n'}} |\psi_{E_l}(\vec{n'})|^2=O(1)\} \ .
\end{gather}

The expression $N^{D_q}$ characterizes the scaling of the number of sites, from which the dominant contribution to $I_q$ is given.
So, for the localized states $D_{q>0}=0$, while in the ergodic phase $D_q = 1$. The non-ergodic extended states, characterized by fractional $0<D_q<1$, are called fractal (if $D_{q>1/2} = D$) or multifractal (if $D_q > D_{q+1}$).
Next, we will provide the estimates for the fractal dimensions $D_2$ and $D_\infty$. Hence, we count the number of resonances at a certain distance $R$ where $\delta_R\leq J_R$.

The sites $\vec{m+R}$ being resonant to $\vec{m}$, have the energies $\epsilon_{\vec{m+R}}$, in the interval $J_R$ around $\ep_{\vec{m}}$. On the other hand, the mean level spacing of $R^d$ levels is given by $\delta_R$.
So, for a smooth density of states (DOS), the number of resonances is given by
\begin{gather}\label{eq:Nres}
N_{res,R} \simeq \min\left(\frac{J_R}{\delta_R}, R^d\right) \ .
\end{gather}
The later case of all the resonant sites occurs when $J_R$ is larger than the energy bandwidth $W_R\sim \sqrt{\mean{(\epsilon_{\vec{m+R}}-\epsilon_{\vec{m}})^2}}$.

%\item
With the $D_\infty$-definition, Eq.~\eqref{eq:D_infty}, we estimate it, using the Ioffe-Regel criterion, similarly to the random-banded matrix case~\cite{Wilkinson_1991}.
Indeed, as up to a certain distance $R_\infty$ all the sites are resonant, given by the condition $N_{res, R<R_\infty} = R^d$ ($J_R>W_R$), one can approximate the Hamiltonian as a banded matrix ensemble with the bandwidth $R_\infty^d$, see Appendix~\ref{App_Sec:Banded_mat}. %\cite{SM}.
From~\cite{Wilkinson_1991}, it is known that in this case, the wave function has an extensive localization length $\xi$, which determines the fractal dimension $D_\infty$:
% CAN BE SHORTENED BY PUTTING INLINE!!
\begin{gather}\label{eq:D_inf_res}
N^{D_\infty}\sim \xi \sim R_\infty^{2d}  \ .
\end{gather}

%\item
The fractal dimension $D_2$ of the standard IPR can be estimated in two ways.
First, from the PLRBM resonance counting. In this case $N^{D_2}$ is given by a volume $R_*^d$ at the distance $R_*$, below which at least one resonance is present, i.e. %(see Eq.~\eqref{eq:R_*})
\begin{gather}\label{eq:D2_PLRBM}
N^{D_2} \simeq R_*^d
\end{gather}
If there is no such $R_*$ or $R_\infty$, satisfying the above conditions, and instead one sees the resonances at $R>R^{**}$, also given by Eq.~\eqref{eq:R*}, then the system is truly long-ranged and is similar to the RP model~\cite{RP,Kravtsov_NJP2015}.
In the latter case, the fractal dimensions $D_{q>1/2}=D$ are usually equal to each other and can be found %either
by counting all the resonances
\begin{gather}\label{eq:D_RP_resonances}
\max\nolimits_R N_{res,R} \sim N^{D/2} \ .
\end{gather}
%or using the Fermi's Golden rule via the scaling of ratio
%\begin{gather}\label{eq:RP_FGR}
%N^{D} \sim \frac{\Gamma}{\delta_N}
%\end{gather}
%of the level broadening $\Gamma$ to the mean level spacing $\delta_N$:
%\begin{gather}
%\Gamma = \frac{2\pi}{\hbar} \sum_n \rho(E)|H_{\vec{mn}}|^2 \ , \ \rho(E) \sim 1/(N \delta_N) \ .
%\end{gather}

%\item
Second, to take into account the change of the resonance structure after hybridization (absent in the above counting), we consider only the hopping terms at the shortest distance $R=1$, $J_R \to J_1 \delta_{R,1}$, and map the system to the $d_{\rm{eff}}$-dimensional Anderson model, where $d_{\rm{eff}}$ is the power of the decay of $\delta_R$ with $R$, see Appendix~\ref{App_Sec:ALT_d_eff} %\cite{SM}
\begin{gather}\label{eq:delta_deff}
\delta_R = \frac{W_{\rm{eff}}(N)}{R^{d_{\rm{eff}}}} \ .
\end{gather}
The number of sites in such a $d_{\rm{eff}}$-dimensional model grows as
%\begin{gather}
$N_{\rm{eff}}(R) \sim R^{d_{\rm{eff}}}$.
%\end{gather}
If the ratio of the effective disorder $W_{\rm{eff}}(N)/J_1$ decays with $N$, like in~\cite{Das2023beta-ens_IMK}, the fractal dimension support $N^{D_2}$ is given by the number of sites $N_{\rm{eff}}(\xi_1)$ at the distance of the localization length $\xi_1$.
The latter is estimated for $d_{\rm{eff}}\leq 1$ from its $1$d expression~\cite{Izrailev1998,Sanchez1,Das2023beta-ens_IMK},
%\begin{gather}
$\xi_1(N) \sim \left(J_1/W_{\rm{eff}}\right)^2$,
%\end{gather}
leading to (Appendix~\ref{App_Sec:ALT_d_eff})%~\cite{SM}
\begin{gather}\label{eq:D2_1d_ALT}
N^{D_2} \sim N_{\rm{eff}}(\xi_1) = \left(J_1/W_{\rm{eff}}\right)^{2d_{\rm{eff}}} \ .
\end{gather}
Note that the effective Anderson model, with only $J_{R_0}$ at fixed $R=R_0$ taken into account, gives even smaller estimates
%\begin{gather}
$\xi_{R_0}\sim {R_0}\left(J_{R_0}/W_{\rm{eff}}\right)^2 \simeq \xi_1 {R_0}^{-(2a-1)} \ll \xi_1$.

To sum up, $N^{D_\infty}$ ($N^{D_2}$) is given by the maximum of the above lower bounds Eqs.~\eqref{eq:D_inf_res} (Eqs.~\eqref{eq:D2_PLRBM},~\eqref{eq:D2_1d_ALT}) and~\eqref{eq:D_RP_resonances}.
%\end{gather}
%and
%\begin{gather}
% N_{\rm{eff}}(\xi_R) \simeq  N_{\rm{eff}}(\xi_1) R^{-2d_{\rm{eff}}(2a-1)} \ll N_{\rm{eff}}(\xi_1) \ .
%\end{gather}
%\end{itemize}

\begin{figure*}[t]
\label{Fig2}
    \includegraphics[width=1.0\textwidth]{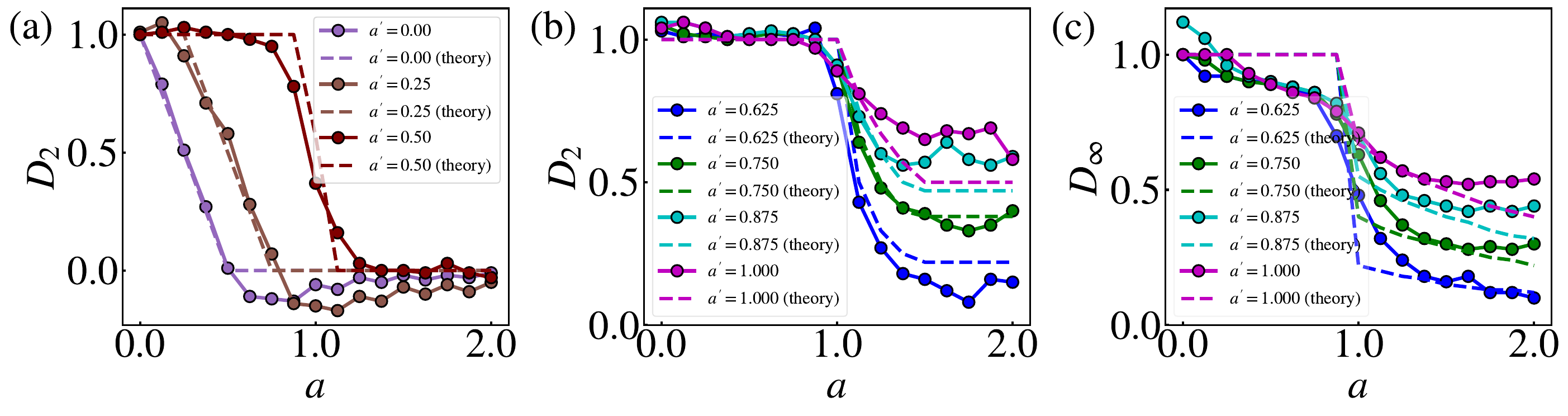}
    \caption{\textbf{Fractal dimensions $D_2$ and $D_\infty$, Eq.~\eqref{eq:IPR_q}}, vs the real-space hopping decay rate $a$ for different values of the momentum-space one $a'$: eigenstate real-space fractal dimension $D_2$ in the (a)~RP-like, $a'<0.5$ and (b)~power-law-like, $a'>0.5$, fractal phases and (c)~$D_\infty$ for $a'>0.5$. Dashed lines in (a-c) are given by Eqs.~\eqref{eq:D_res},~\eqref{eq:D2_res}, and~\eqref{D_inf_res}, respectively.
    Deviations of the data from the analytical predictions are due to the finite-size effects. Measures to take care of these and their extrapolations appear in Appendix~\ref{App_Sec:Dq(N)}.%~\cite{SM}.
    }
\end{figure*}

%%%%%%%%%%%%%%%%%%%%%%%%%%%%%%%%%%%%%%%%%%%%%%%%%%%%%%%%%%%%%%
\textit{Exemplary model --}
%%%%%%%%%%%%%%%%%%%%%%%%%%%%%%%%%%%%%%%%%%%%%%%%%%%%%%%%%%%%%%
To demonstrate the applicability of the above method, now we focus on the $1$d model~(\ref{eq:ham},~\ref{eq:J_p}), with the i.i.d. translation-invariant $J_{m,n} \equiv J_{m-n}$. The model is comprised of the power-law decaying hopping and correlated on-site potential~\eqref{eq:eps_n},~\eqref{eq:J_p}, where the correlator is given by,
\begin{equation}\label{eq:<eps_n eps_m>}
\langle\epsilon_{\vec{n}} \epsilon_{\vec{m}}\rangle = %\sum_{p,q=1}^{N-1} \frac{\langle \bar{j}_{p} \bar{j}_{q}^*\rangle }{|p|^{a^{\prime}}|q|^{a^{\prime}}} e^{{i\frac{2\pi}{N}}(pn-qm)} =
\sum_{p=1}^{N-1} \frac{e^{{i\frac{2\pi}{N}}p(n-m)}}{|p|^{2 a^{\prime}}} \sim \left(\frac{|m-n|}{N}\right)^{2a^\prime-1} + O(1) \ .
\end{equation}
This model is self-dual to itself in the momentum space for $a=a'$. Thus, further, we focus on the coordinate-basis phase diagram.

Among limiting cases: the case of $a'=0$ corresponds to the PLRBM~\cite{Levitov1989,PLRBM,Nosov2019correlation}; $a=a'=0$ gives TI RP at its self-dual point ($\gamma=1$)~\cite{Nosov2019correlation}, while $a, a'\to\infty$ brings us to the short-range model with deterministic potential, similar to AA~\cite{AA}.
In addition, at all $a^\prime>1$, the corresponding series~\eqref{eq:eps_n} is absolutely convergent, i.e., the potential is nearly deterministic. Thus, we do not expect any effect of this disorder potential on the localization properties.
Thus, later we focus mostly on $a^\prime<1$.

In the saddle-point approximation in $N$ and $R$, the main contributions to the $R$-dependent bandwidth $W_R$, determined via~\eqref{eq:<eps_n eps_m>}, is given by the following momenta $p\sim O(1)$, $p\sim O(N/R)$, and $p \sim O(N)$, see Appendix~\ref{App_Sec:delta_R_correlated} %Sec.~II in~\cite{SM}
\begin{multline}\label{eq:<(eps_n - eps_m)^2>_res}
\langle(\epsilon_{\vec{m+R}}-\epsilon_{\vec{m}})^2\rangle =
\sum_{p=1}^{N-1} \frac{4 \sin^2\left(\frac{\pi}{N}p R\right)}{|p|^{2 a^{\prime}}} \sim \\ \sim \left(\frac{R}{N}\right)^2 + N^{1-2a^{\prime}} + \left(\frac{R}{N}\right)^{2a^{\prime}-1}
\ .
\end{multline}

For $a'<1/2$ the main contribution %to~\eqref{eq:<(eps_n - eps_m)^2>_res}
is given by $p\sim O(N)$, i.e. by $\langle\epsilon_{\vec{n}}^2\rangle + \langle \epsilon_{\vec{m}}^2\rangle$. This is equivalent to the i.i.d. $\epsilon_{\vec{n}}$ case.
In the rest interval in focus, $1/2<a'<1$, the dominant contribution is from $p\sim O(N/R)$, giving the $R$-dependent mean level spacing
\begin{gather}\label{eq:delta_R_res}
\delta_R = \left\{
             \begin{array}{ll}
               {N^{1/2-a^\prime}}/{R}, & a^\prime<1/2 \\
               {N^{-(a^\prime-1/2)}}/{R^{3/2-a^\prime}}, & a^\prime>1/2
             \end{array}
           \right. \ .
\end{gather}
This expression opens the way to determine the phase diagram in the entire parameter range.

\begin{figure}[b!]
\label{Fig3}
    \includegraphics[width=1\columnwidth]{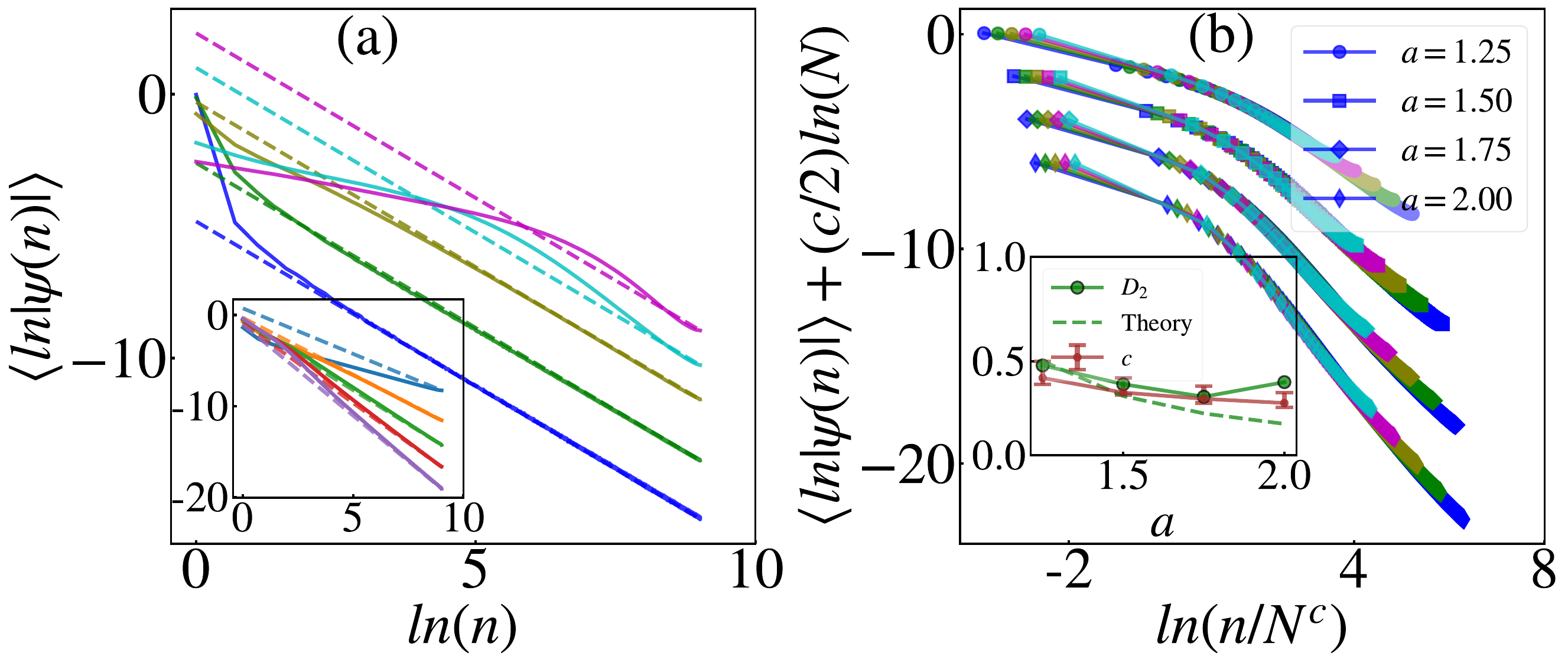}
    \caption{\textbf{Spatial wave-function decay} in
    (a)~the power-law localized phase at $a=1.25$ for different $a'$ from $0$ to $1$ with the step $0.25$ (bottom to top), with dashed lines given by Eq.~\eqref{eq:WF_decay}. (inset)~for different $a$ from $1$ to $2$ with the step $0.25$ (top to bottom) at $a'=0.5$.
    (b)~in the fractal phase $a'=0.75$ for different $a$, with the collapse of the wave-function ``head''. (inset)~the extracted collapse parameter $c$ vs $a$, compared to the fractal dimension $D_2$ from Fig.~\ref{Fig2}(b), and its analytical expression, Eq.~\eqref{eq:D2_res}. %In all panels dashed lines show analytical predictions.
    }
\end{figure}

%%%%%%%%%%%%%%%%%%%%%%%%%%%%%%%%%%%%%%%%%%%%%%%%%%%%%%%%%%%%%%
\textit{Localized phase --}
%%%%%%%%%%%%%%%%%%%%%%%%%%%%%%%%%%%%%%%%%%%%%%%%%%%%%%%%%%%%%%
The localized phase occurs, when $\delta_R>J_R$ for all $R>O(1)$, i.e., at
\begin{gather}
a^\prime\leq1/2 \text{ and }
a > a^\prime + 1/2 \ ,
\end{gather}
see Fig.~\ref{Fig2}(a) and red lines in Fig.~\ref{Fig1}.
Indeed, due to the prefactor $N^{1/2-a^\prime}$, the level spacing is enhanced at $a^\prime<1/2$ with respect to the uncorrelated case $a^\prime>1/2$. This leads to the shift of the Anderson localization transition to the larger $a$-values with respect to $a=1$ in the PLRBM case. The eigenstate spatial decay, $\mean{\ln |\psi_E(\vec{n})|}$ with respect to the distance $n$ from the wave-function maximum~\cite{Deng2018Duality,Nosov2019correlation,Kutlin2020_PLE-RG,Motamarri2022RDM,Tang2022nonergodic,Deng2022AnisBM} is given by a perturbative term of the power-law decaying hopping term, normalized to the bandwidth, see Fig.~\ref{Fig3}(a)
\be\label{eq:WF_decay}
|\psi_E({\vec{n+R}})|\simeq J_R/W_R \sim N^{-(1/2-a')}/R^a \ .
\ee

On the other hand, the level-spacing reduction, due to the correlations at $a^{\prime}>1/2$, delocalizes the system up to
\begin{gather}\label{eq:R_*}
R_* = N^{\frac{a^\prime-1/2}{a + a^\prime - 3/2}} \ ,
\end{gather}
as at all $R<R_*$ we have $\delta_R<J_R$. Note that at $a<1$, $R_*$ scales faster than $N$. Therefore the eigenstates at $a<1$, $a^\prime>1/2$ are expected to be ergodic. This is also the case for any $a$ at $a'>1$, see Fig.~\ref{Fig2} and the blue lines in Fig.~\ref{Fig1}(b).

%%%%%%%%%%%%%%%%%%%%%%%%%%%%%%%%%%%%%%%%%%%%%%%%%%%%%%%%%%%%%%
\textit{Fractal phases --}
%%%%%%%%%%%%%%%%%%%%%%%%%%%%%%%%%%%%%%%%%%%%%%%%%%%%%%%%%%%%%%
From the previous consideration, there are only two parameter intervals left for the fractal phases: one in the uncorrelated part,
$a^\prime<1/2$, $a<a^\prime+1/2$, and another in the correlated one, $a>1$, $1/2<a^\prime< 1$.

In the uncorrelated case, $a^\prime<1/2$, $0<a<a^\prime+1/2\leq 1$ from Eqs.~\eqref{eq:delta_R_res} and~\eqref{eq:Nres}
\begin{gather}
N_{res,R} \simeq \min\left(\frac{R^{1-a}}{N^{a^\prime-1/2}}, R^1\right) \ .
\end{gather}
We see that no $R_*$, Eq.~\eqref{eq:R*}, can be found, and thus, the fractal dimensions are given by~\eqref{eq:D_RP_resonances}
\begin{gather}\label{eq:D_res}
%\max_R N_{res,R} = N_{res,N} \sim N^{1/2-a-a^\prime} \Leftrightarrow
D = 1/2-a-a^\prime
\end{gather}
in agreement with the numerics in Fig.~\ref{Fig2}(a).

In the correlated case $1/2<a^\prime< 1$, $a>1$ from Eq.~\eqref{eq:Nres}
\begin{gather}\label{eq:N_res_R_infty}
N_{res,R} \simeq \min\left(\frac{N^{a^\prime-1/2}}{R^{a^\prime+a-3/2}}, R^1\right) \Rightarrow  R_\infty = N^{\frac{a^\prime-1/2}{a + a^\prime - 1/2}} \ ,
\end{gather}
$R_*$ is given by~\eqref{eq:R_*} and from Eqs.~\eqref{eq:delta_deff}-\eqref{eq:D2_1d_ALT}
\begin{gather}\label{eq:xi1_1d_ALT}
\xi_1(N)\sim N^{-(a^\prime-1/2)} \ , \ d_{\rm{eff}}=3/2-a^\prime . %\Leftrightarrow D_2\geq 2(a^\prime-1/2)(3/2-a^\prime).
\end{gather}
Thus, from~\eqref{eq:D_inf_res} and~\eqref{eq:N_res_R_infty}
\begin{gather}\label{D_inf_res}
D_\infty = \frac{\ln \lrp{R_\infty^2}}{\ln N}=\frac{2a^\prime-1}{a + a^\prime - 1/2} < D_2 \ ,
\end{gather}
while from Eqs.~\eqref{eq:D2_PLRBM},~\eqref{eq:D2_1d_ALT},~\eqref{eq:R_*}, and~\eqref{eq:xi1_1d_ALT}
\begin{gather}\label{eq:D2_res}
D_2= (a^\prime-1/2)\max\left(\frac{1}{a + a^\prime - 3/2},3/2-a^\prime\right) \ ,
\end{gather}
that both also well agree with the numerics, Fig.~\ref{Fig2}(b, c).
The above result, Eq.~\eqref{eq:D2_res}, is also consistent with the collapse of the wave-function head $R<R^*$, see Fig.~\ref{Fig3}(b).

In addition, numerically, in Fig.~\ref{Fig1}(a) we have checked the spectral properties via the $r$-statistics of levels $E_n$, which is given by,
\be\label{eq:r-stat}
r = \min\lrp{\tilde r_n, \frac1{\tilde r_n}} \text{, where } \tilde r_n=\frac{E_{n+1}-E_n}{E_n-E_{n-1}} \ .
\ee
While in the ergodic (localized) phases, $r$ takes Gaussian Unitary (Poisson) value $0.599$ ($0.386$), the behavior in fractal phases is different. Indeed, in the uncorrelated one, $a^\prime<1/2$, the behavior is similar to RP model~\cite{Kravtsov_NJP2015}, while the correlated phase, $a^\prime>1/2$, shows Poisson, similarly to~\cite{Tang2022nonergodic,Das2023beta-ens_IMK}, see Appendix~\ref{App_Sec:Tang} %Sec. III in~\cite{SM} 
for more details.

%%%%%%%%%%%%%%%%%%%%%%%%%%%%%%%%%%%%%%%%%%%%%%%%%%%%%%%%%%%%%%
\textit{Conclusions and outlook —-}
%%%%%%%%%%%%%%%%%%%%%%%%%%%%%%%%%%%%%%%%%%%%%%%%%%%%%%%%%%%%%%
To sum up, the main result of this paper is the development of the approach given by Eqs.~\eqref{eq:delta_R}-\eqref{eq:eps_n},~\eqref{eq:Nres}-\eqref{eq:D2_1d_ALT}, generic for the description of both the localization and ergodic transitions, as well as the fractal dimensions in the long-range models with the correlated on-site disorder. This technique works well both for the range of known models, including PLRBM~\cite{PLRBM}, RP~\cite{Kravtsov_NJP2015}, TI RP~\cite{Nosov2019correlation}, $\beta$-ensemble~\cite{Das2022beta-ens,Das2023beta-ens_IMK} and many others~\cite{Tang2022nonergodic,Deng2019Quasicrystals,Gopalakrishnan2017}, as well as for a new class of self-dual models, considered here.

This approach is also might be applicable to the models with Wannier-Stark mechanism of localization~\cite{Schulz2019Stark_MBL,Morong2021observation,Dwiputra2022Stark-Mosaic,gao2023observation} and to the ones with the non-Hermitian~\cite{DeTomasi2022nH-RP,DeTomasi2023nH-PLRBM,DeTomasi2022nonHerm_MBL} or fractal~\cite{sarkar2023fract-RP} distribution of the on-site potential.

It would be of particular interest and high demand to generalize this approach beyond the translation invariant on-site disorder correlations $\mean{\epsilon_{\vec{m}}\epsilon_{\vec{n}}}\ne F(\vec{m-n})$ as it is relevant for non-lattice models, such as the Anderson model on the hierarchical graphs and the many-body localization phenomenon from the Hilbert-space perspective.
Another interesting direction of generalization is given by the set of models with complete~\cite{Burin1989,Ossipov2013,Deng2018Duality,Nosov2019correlation,Kutlin2020_PLE-RG,Motamarri2022RDM} or partial~\cite{Nosov2019mixtures,Kutlin2021emergent} correlations in off-diagonal terms, where the renormalization group is applicable only after a certain matrix-inversion trick~\cite{Nosov2019correlation}.

%%%%%%%%%%%%%%%%%%%%%%%%%%%%%%%%%%%%%%%%%%%%%%%%%%%%%%%%%%%%%%
\begin{acknowledgements}
We thank G. De Tomasi and V.~E.~Kravtsov for illuminating discussions.
I.~M.~K. acknowledges the support from
Russian Science Foundation (Grant No. 21-12-00409).
%the European Research Council under the European Union's Seventh Framework Program Synergy ERC-2018-SyG HERO-810451.
\end{acknowledgements}
%%%%%%%%%%%%%%%%%%%%%%%%%%%%%%%%%%%%%%%%%%%%%%%%%%%%%%%%%%%%%%
\bibliography{Lib}

\appendix

\begin{widetext}
%%%%%%%%%%%%%%%%%%%%%%%%%%%%%%%%%%%%%%%%%
\section{General translation-invariant correlations of diagonal elements $\epsilon_{\vec{n}}$}\label{App_Sec:general_correlations}
%%%%%%%%%%%%%%%%%%%%%%%%%%%%%%%%%%%%%%%%%

Let's consider generally correlated $\epsilon_{\vec{n}}$ using the first part of Eq.~\eqref{eq:eps_n} %(5)
\begin{equation}
\epsilon_{\vec{n}}=\sum_{\vec{p}} J_{\vec{p}} e^{{i\frac{2\pi}{N}}\vec{p n}} \ ,
\end{equation}
with a generic correlation function in the momentum space
\begin{gather}
\langle J_{\vec{p}} J_{\vec{q}}^* \rangle = F\left(\vec{p}_+=\frac{\vec{p+q}}{2},\vec{p}_-=\vec{p-q}\right) \ .
\end{gather}

The correlator $\langle \epsilon_{\vec{n}} \epsilon_{\vec{m}} \rangle$ in this case is given by
\begin{gather}
\langle \epsilon_{\vec{n}} \epsilon_{\vec{m}} \rangle = \sum_{\vec{p,q}\neq 0} F\left(\vec{p}_+=\frac{\vec{p+q}}{2},\vec{p}_-=\vec{p-q}\right) e^{{i\frac{2\pi}{N}}(\vec{p n-q m})}
\end{gather}

Using the fact that $\epsilon_{\vec{n}}$ is real
\begin{gather}
\epsilon_{\vec{n}} = \epsilon_{\vec{n}}^* \quad\Leftrightarrow \quad J_{\vec{p}} = J_{-\vec{p}}^*
\end{gather}
and the symmetry of the correlator
\begin{gather}
%\langle J_p J_q^* \rangle = \langle J_q^* J_p \rangle \ , \quad
\langle J_{\vec{p}} J_{\vec{q}}^* \rangle = \langle J_{-\vec{q}}J_{-\vec{p}}^* \rangle
\end{gather}
one puts the following restriction on the function $F(\vec{p}_+,\vec{p}_-)$
\begin{gather}\label{eq:F(p+,p-)}
%F(p_+,-p_-) = F^*(p_+,p_-) \ , \quad
F(-\vec{p}_+,\vec{p}_-) = F(\vec{p}_+,\vec{p}_-) \ .
\end{gather}

In addition, if one assumes the correlator $\langle \epsilon_{\vec{n}} \epsilon_{\vec{m}} \rangle$ to be translation-invariant
\begin{gather}\label{eq:<eps_n eps_m>_TI}
\langle \epsilon_{\vec{n}} \epsilon_{\vec{m}} \rangle = \langle \epsilon_{\vec{n}+s} \epsilon_{\vec{m}+s} \rangle \equiv \frac{1}{N}\sum_s \langle \epsilon_{\vec{n}+s} \epsilon_{\vec{m}+s} \rangle \ ,
\end{gather}
this will lead to
\begin{multline}
F\left(\vec{p}_+=\frac{\vec{p+q}}{2},\vec{p}_-=\vec{p-q}\right) =\frac{1}{N^2}\sum_{\vec{n,m}}\langle \epsilon_{\vec{n}} \epsilon_{\vec{m}} \rangle e^{-{i\frac{2\pi}{N}}(\vec{p n-q m})}=\\
=\frac{1}{N^3}\sum_{\vec{n,m,s}}\langle \epsilon_{\vec{n+s}} \epsilon_{\vec{m+s}} \rangle e^{-{i\frac{2\pi}{N}}(\vec{p n-q m + s(p-q)})}=\\
=\frac{1}{N^2}\sum_{\vec{n,m}}\langle \epsilon_{\vec{n}} \epsilon_{\vec{m}} \rangle e^{-{i\frac{2\pi}{N}}(\vec{p n-q m} )}\sum_{\vec{s}} \frac{e^{-{i\frac{2\pi}{N}}\vec{s(p-q)}}}{N} =
\delta_{\vec{p}_-,0} F\left(\vec{p}_+=\vec{p}=\vec{q}\right) \ ,
\end{multline}
where according to~\eqref{eq:F(p+,p-)} the function $F(\vec{p}_+)$ is even in $\vec{p}_+$.

This confirms that a generic set of correlated $\epsilon_{\vec{n}}$ with a translation-invariant correlator~\eqref{eq:<eps_n eps_m>_TI} is given by uncorrelated $J_{\vec{p}}$ with a certain spectrum of the variance
\begin{gather}
\langle J_{\vec{p}} J_{\vec{q}}^* \rangle = \delta_{\vec{p,q}} F(\vec{p}_+=\vec{p}=\vec{q}) \ .
\end{gather}

%%%%%%%%%%%%%%%%%%%%%%%%%%%%%%%%%%%%%%%%%
\section{Contributions to the level difference $\epsilon_{\vec{n}}-\epsilon_{\vec{m}}$}\label{App_Sec:delta_R_correlated}
%%%%%%%%%%%%%%%%%%%%%%%%%%%%%%%%%%%%%%%%%
In order to obtain Eq.~\eqref{eq:<(eps_n - eps_m)^2>_res} %(16)
in $1$d, here we consider the contributions to $\epsilon_\vec{m+R}-\epsilon_\vec{m}$ in more details
\begin{gather}
\epsilon_\vec{m+R}-\epsilon_\vec{m} =
\sum_{p=1}^{N-1} \frac{2 i j_p e^{-i\frac{2\pi}{N}p(m+R/2)}}{|p|^{a^{\prime}}} \sin \left(\frac{\pi}{N}p R\right)\equiv
\sum_{p=1}^{N-1} \frac{2 \tilde j_p}{|p|^{a^{\prime}}} \sin \left(\frac{\pi}{N}p R\right)
\ ,
\end{gather}
where $\tilde j_p = i j_p e^{-i\frac{2\pi}{N}p(m+R/2)}$ is a set of i.i.d. complex variables (with random homogeneously distributed phases).
From this one can immediately see the random-sign series, which absolutely converges at $a^\prime>1$ at few first terms $p\sim O(1)$ as well as the corresponding $\epsilon_{\vec{m}}$.
In this situation, the diagonal potential is not anymore random and can be approximated by the first terms $p=\pm 1$
\begin{gather}
\epsilon_{n}=|J_{p=1}|\cos\left(\frac{2\pi n}{N}+{\rm arg} J_{p=1}\right) \ .
\end{gather}
Therefore further in this Appendix we consider $a^\prime\leq 1$ where the randomness of $\epsilon_{\vec{n}}$ is given by the contribution from the extensive number of terms in the series.

To consider $a^\prime<1$ let's first fix $m$ and consider the ranges of $p_k\leq p<2p_k$ with $p_k$ summands in each, $p_{k+1}=2p_k$, and find from which one the maximal contribution to the sum is expected.
At a certain $p_k$ the contribution is given by the series with a certain amplitude $A_k$ (smoothly dependent on $p$ in $p_k$-range) and a random sign.
Therefore the $p_k$-contribution $(\epsilon_{\vec{m+R}}-\epsilon_{\vec{m}})_k$ is given by $\sim A_k p_k^{1/2}$.
\begin{itemize}
  \item For $p_k\ll N/R$
\begin{gather}
A_p = \frac{\left|\sin \left(\frac{\pi}{N}p R\right)\right|}{|p|^{a^\prime}} \simeq \frac{\pi |p|^{1-a^\prime} R}{N} \ ,
\end{gather}
i.e. the contribution
\begin{gather}
(\epsilon_{\vec{m+R}}-\epsilon_{\vec{m}})_k \simeq \frac{\pi |p|^{3/2-a^\prime} R}{N}
\end{gather}
grows with $p$.

  \item For $p_k\gg N/R$
\begin{gather}
A_p = \frac{1}{|p|^{a^\prime}} \ ,
\end{gather}
while the random variable is given by
\begin{gather}
\bar j_p = i j_p e^{-i\frac{2\pi}{N}p(m+R/2)}\sin \left(\frac{\pi}{N}p R\right)
\end{gather}
i.e. the contribution
\begin{gather}
(\epsilon_\vec{m+R}-\epsilon_\vec{m})_k \simeq |p|^{-(a^\prime-1/2)}
\end{gather}
decays with $p$ at $a^\prime>1/2$ and grows with $p$ at $a^\prime<1/2$.
\end{itemize}
To sum up, at $1/2<a^\prime<1$ the maximal contribution is given by $p_k \simeq N/R$ and takes the form
\begin{gather}
\epsilon_{m+R}-\epsilon_{m} \sim \left(\frac{R}{N}\right)^{a^\prime-1/2} \ .
\end{gather}

On the other hand, at $a^\prime<1/2$ the main contribution is given by $p_k\sim N$ and takes the form
\begin{gather}
\epsilon_{m+R}-\epsilon_{m} \sim 2\epsilon_{m} \sim N^{1/2-a^\prime} \ .
\end{gather}
This means that $\epsilon_{m}$ look like i.i.d. random numbers of the above amplitude.

One can see it directly from Eq.~(15) %\eqref{eq:<eps_n eps_m>}
\begin{equation}
\langle\epsilon_{n} \epsilon_{m}\rangle = \sum_{p=1}^{N-1} \frac{e^{{i\frac{2\pi}{N}}p(n-m)}}{|p|^{2 a^{\prime}}} \ .
\end{equation}
Indeed, providing the same analysis of $p_k$-contributions, one can see that at $p_k\ll N/R$ the contribution does not have a random sign and grows with $p$
\begin{gather}
A_k \sim \frac{1}{|p|^{2 a^{\prime}}} \quad \Rightarrow \quad \langle\epsilon_{n} \epsilon_{m}\rangle\sim A_k p_k \sim |p_k|^{1-2a^\prime} \ ,
\end{gather}
while at $p_k\gg N/R$ it becomes sign-alternating and gives the contribution
\begin{gather}
A_k \sim \frac{1}{|p|^{2 a^{\prime}}} \quad \Rightarrow \quad \langle\epsilon_{n} \epsilon_{m}\rangle\sim A_k p_k^{1/2} \sim |p_k|^{1/2-2a^\prime} \ ,
\end{gather}
decaying with $p_k$ at $a^\prime>1/4$ and growing otherwise.

At $1/4<a^\prime<1/2$ the correlator is given by $p_k\sim N/R$
\begin{gather}
\langle\epsilon_{n} \epsilon_{m}\rangle \sim \left(\frac{N}{R}\right)^{1-2a^\prime} \sim \frac{\langle\epsilon_{n}^2\rangle}{R^{1-2a^\prime}} \ll\langle\epsilon_{n}^2\rangle \ ,
\end{gather}
while at $a^\prime<1/4$
\begin{gather}
\langle\epsilon_{n} \epsilon_{m}\rangle \sim N^{1/2-2a^\prime} \sim \frac{\langle\epsilon_{n}^2\rangle}{N^{1/2}} \ll\langle\epsilon_{n}^2\rangle \ .
\end{gather}
In both cases the correlations between $\epsilon_{n}$ and $\epsilon_{m}$ are parametrically smaller than the amplitudes of these terms.

%%%%%%%%%%%%%%%%%%%%%%%%%%%%%%%%%%%%%%%%%
\section{Finite-size effect analysis of fractal dimensions}\label{App_Sec:Dq(N)}
%%%%%%%%%%%%%%%%%%%%%%%%%%%%%%%%%%%%%%%%%
In the main text for the exemplary model, given by Eqs.~(1, 5, 6, 15), the numerical calculation of fractal dimensions $D_2$ and $D_\infty$ encoded in Eq.~(7), show rather strong finite-size effects (see Fig.~1).

As a remedy, we have considered the standard extrapolation scheme~\cite{EversMirlin2008,DeLuca2014,Kravtsov_NJP2015,Deng2018Duality,Nosov2019correlation,Nosov2019mixtures,Motamarri2022RDM} for the fractal dimensions.

\begin{figure}[th]
  \center{
  \includegraphics[width=\columnwidth]{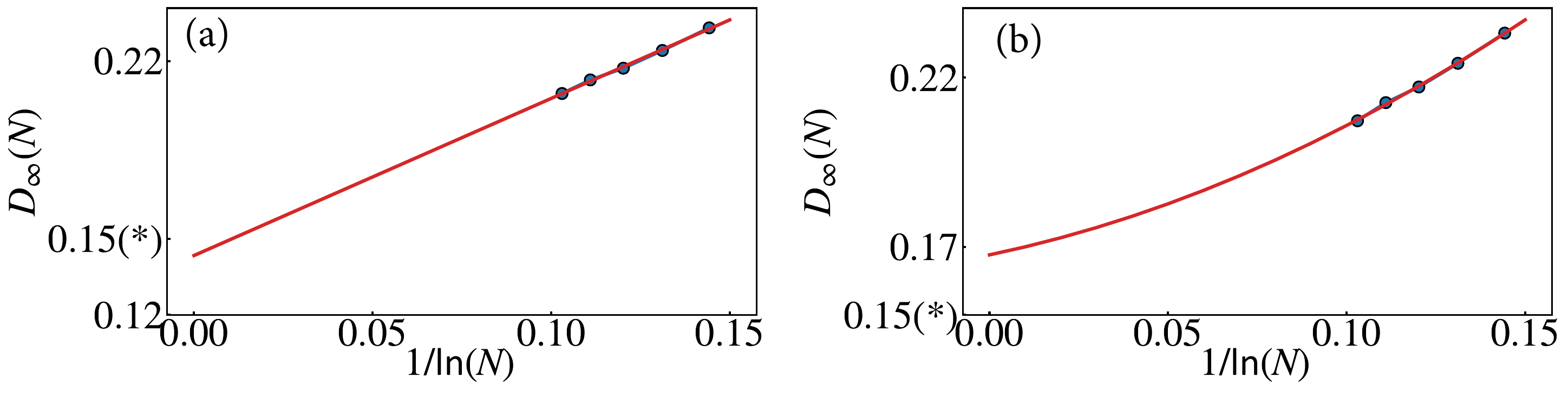}
  \hfill
  \hfill
  \includegraphics[width=\columnwidth]{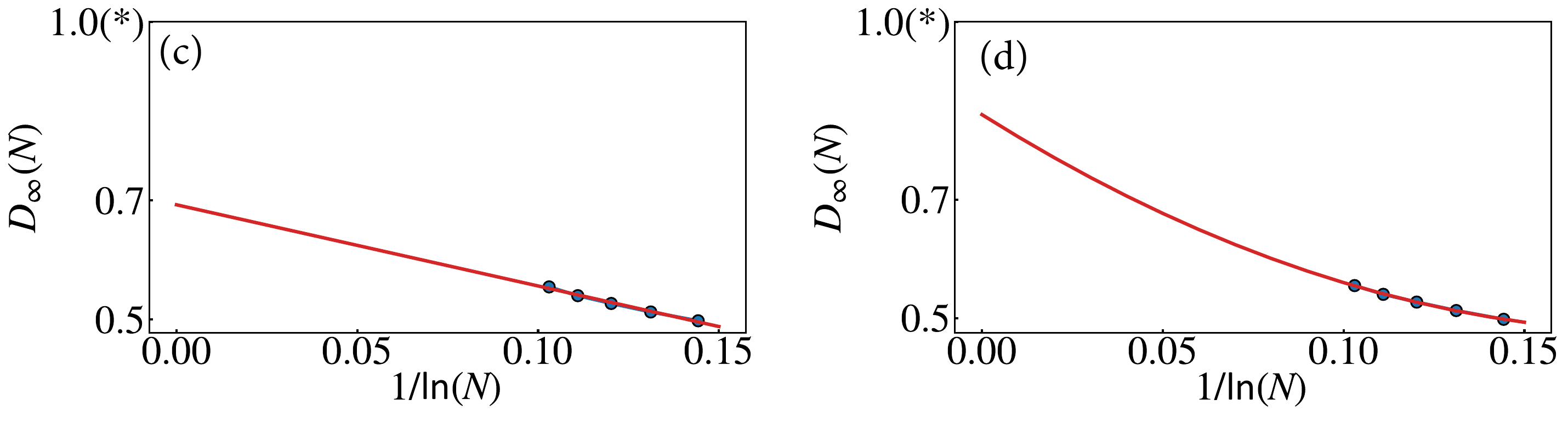}
  }
  \caption{\textbf{Finite-size extrapolation of the fractal dimension } $D_{\infty}$
  for the parameters with (a, b)~good $a=1.5$, $a'=0.625$ and (c, d)~bad $a=0.875$, $a'=0.625$ extrapolation. In order to show the difference (a,c)~in the left column we show the standard linear extrapolation in $1/\ln N$, while (b,d)~in the right column~--~the quadratic one in $1/\ln N$. The theoretically predicted point for the extrapolation are shown by the asterisk on the vertical axis.
  %(a)~$a= 0.50, ~a'=0.25$, (b)~$a= 0.50, ~a'=1.00$, and (c)~$a= 1.50, ~a'=1.00$ in the upper panel and (a)~$a= 1.00, ~a'=0.25$, (b)~$a= 1.75, ~a'=1.00$, and (c)~$a= 0.25, ~a'=1.00$ in the lower panel.
  The system sizes taken for the calculations are $N=1024,~2048,~4096,~8192$, and $16384$.
  }
  \label{Fig:D2_extr}
\end{figure}

The finite-size fractal dimension is defined by the formula $D_q(N) = \ln I_q/\lrb{(1-q)\ln N}$, with the generalized inverse participation ratio (IPR) of the order $q$, defined via,
\be\label{App:ipr}
I_q = \sum_i |\psi_n(i)|^{2q} \ .
\ee
In order to avoid the parasitic contributions from measure zero of the special eigenstates, we focus on the typical averaging of the IPR both over disorder and eigenstates implemented by,
\be\label{App:ipr_typ}
I_{q,typ} = e^{\mean{\ln I_q}} = c_q N^{(1-q)D_{q,typ}}. \
\ee
Later on, we shall omit the subscript ``typ'' for brevity.

As the main contribution to IPR is given by the scaling exponent $D_q$ and the prefactor $c_q$ in~\eqref{App:ipr_typ}, one obtains
\be\label{eq:Dq(a,N)}
D_q(N) = D_q + \frac{(1-q)^{-1}\ln c_q}{\ln N} \ .
\ee

The extrapolation of $D_q(N)$ vs $1/\ln N$ extracted from $I_q$ is shown in Fig.~\ref{Fig:D2_extr}. One can see that for the parameters with good extrapolated values (panels (a,~b) of Fig.~\ref{Fig:D2_extr}), already the linear extrapolation gives an accurate result, and the nonlinear corrections cannot improve it.
At the same time, the parameters where the extrapolation is not so good (panels~(c,~d) in Fig.~\ref{Fig:D2_extr}) the curvature in data can significantly improve the extrapolation if taken into account via the nonlinearity of $D_q(N)$ vs $1/\ln N$. Such kinds of the finite-size effects lead to the deviations of the extrapolated values in Fig.~2 of the main text.

%%%%%%%%%%%%%%%%%%%%%%%%%%%%%%%%%%%%%%%%%
\section{Discussion of the long-range correlated models}\label{App_Sec:BM_finite_W_ALT}
%%%%%%%%%%%%%%%%%%%%%%%%%%%%%%%%%%%%%%%%%
Starting from a seminal paper Ref.~\cite{Burin1989}, people have been considering the set of models with fully-correlated (deterministic) off-diagonal elements, such as,
\be
J_{m,m+R} \sim \frac{1}{R^a} \
\ee
and certain non-random (in principle, angle-dependent) prefactor.

As has been shown in~\cite{Malyshev2000,Malyshev2004,Malyshev2005} for $a>d$ and later for $a<d$ in~\cite{Deng2018Duality,Nosov2019correlation,Nosov2019mixtures,Deng2022AnisBM} and many others in such models with uncorrelated diagonal disorder, there is an extensive number, but zero fraction of extended ergodic states at the spectral edge.
For $a>d$, such ergodic states disappear at a certain finite value of the disorder amplitude~\cite{Malyshev2000,Malyshev2004,Malyshev2005}, unlike the case of the PLRBM model.
Similar effects appear at $a<d$ in the PLRBM model with non-Hermitian complex-valued on-site disorder~\cite{DeTomasi2023nH-PLRBM}.

%%%%%%%%%%%%%%%%%%%%%%%%%%%%%%%%%%%%%%%%%
\section{Mapping of all-resonant models to the random banded matrix model}\label{App_Sec:Banded_mat}
%%%%%%%%%%%%%%%%%%%%%%%%%%%%%%%%%%%%%%%%%
In Eqs.~(9-10) of the main text, we have considered the case of the maximal number of resonances given by $N_{res} = R^d$ up to a certain distance $R<R_\infty$. Here, we provide the mapping of such models with $N_{res} = R^d$ to the banded random matrix model~\cite{Wilkinson_1991}.

Indeed, in the Anderson resonant counting for long-range models (see, e.g.,~\cite{BogomolnyRP2018,Nosov2019correlation,LN-RP_WE} and many others) it appears to be that the hopping terms $J_R>W$, that is, $J_{R}$ is large compared to the amplitude of the diagonal disorder ($W$), provide the same contribution as the ones corresponding to $J_R\simeq W$.
The only effect of the former ones is the change of the global density of states of the model.

Now, if one considers the renormalization group approach of~\cite{Levitov1989,Levitov1990} for finite $R$, like in Eqs.~(2-4, 9) of the main text, both for the banded random-matrix model and for the models with $N_{res} = R^d$ up to a certain distance $R_\infty$, one can immediately see that in both cases all the $R^d$ sites at $R\leq R_\infty$ are in resonance with the initial one, and this happens due to the condition $J_R > W$.

This immediately shows that up to $R\leq R_\infty$, both the models can be mapped onto each other. The only possible difference will be in the global density of states due to the statistics of the scenario for $J_R>W$.
This confirms formula~(10) of the main text for the fractal support set $N^{D_\infty}$.

%%%%%%%%%%%%%%%%%%%%%%%%%%%%%%%%%%%%%%%%%
\section{Mapping of short-range models with correlated translation-invariant disorder to the Anderson model of effective dimensionality}\label{App_Sec:ALT_d_eff}
%%%%%%%%%%%%%%%%%%%%%%%%%%%%%%%%%%%%%%%%%
In this part, we focus on the cut model with the only nearest-neighbor hopping term $J_R \to J_1 \delta_{R,1}$ and the correlated disorder with the following correlations, given by the following level spacing,
\begin{gather}\label{eq:delta_deff}
\delta_R\equiv \frac{ \mean{\lrp{\epsilon_{\vec{n+R}}-\epsilon_{\vec{n}}}^2}}{R^d} = \frac{W_{\rm{eff}}(N)}{R^{d_{\rm{eff}}}} \
\end{gather}
at the distance $R$ from any site.

In the standard Anderson model, the corresponding $R$-dependence of level spacing $\delta_R$ is given by the dimensionality $d$ of the lattice and this crucially affects the presence or absence of the delocalization transition.
Thus, in the analogy to the Anderson model, in the above correlated model, one can define the effective dimensionality $d_{eff}$.

Moreover, even for the $N$-dependent disorder amplitude $W_{eff}(N)$ in the standard Anderson model, one can still use expression~\eqref{eq:D2_1d_ALT}  %(14)
for the fractal dimension via the localization length $\xi_1\sim (J_1/W_{eff}(N))^2$, see, e.g.,~\cite{Das2023beta-ens_IMK}.

The resonance condition, entering the standard Ioffe-Regel criterion, includes the dependence on dimensionality only via the level spacing $\delta_R$ on the distance $R$.
Thus, in the models with correlated on-site disorder, Eq.~\eqref{eq:eps_n}, %(5),
the distance dependence of $\delta_R$ forms an effective dimensionality of the system $d_{eff}$, which does affect the results for the fractal dimension, as $d$ does in the standard Anderson model.

Indeed, the localization length $\xi_1$ for $d_{eff}<1$ is given by the standard formula~\cite{Izrailev1998}, while the fractal support set $N^{D_2}$ is determined by the number of resonant sites $N_{eff}(\xi_1)\sim \xi_1^{d_{eff}}$ in the effective $d_{eff}$ space at the distance $\xi_1$.
This concludes our mapping and confirms the formula~\eqref{eq:D2_1d_ALT} %(14)
from the main text.

%%%%%%%%%%%%%%%%%%%%%%%%%%%%%%%%%%%%%%%%%
\section{Comparison of the correlated case with~\cite{Tang2022nonergodic}}\label{App_Sec:Tang}
%%%%%%%%%%%%%%%%%%%%%%%%%%%%%%%%%%%%%%%%%
The correlated fractal case of $1/2<a^\prime< 1$, $a>1$, with
\begin{gather}
\delta_R = \frac{N^{-(a^\prime-1/2)}}{R^{3/2-a^\prime}} \ , \; J_R \sim \frac{1}{R^a}
\end{gather}
has some similarities with the model in~\cite{Tang2022nonergodic}, where
\begin{gather}
\delta_R = \frac{N^{-\gamma/2}}{R} \ , \; J_R \sim \frac{1}{R^{a_{\rm{eff}}}} \ .
\end{gather}
In terms of the ratio $J_R/\delta_R$, one can map two models to each other, with the parameters
\begin{gather}
\gamma = 2 a^\prime-1 \in (0,1) \ , \; a_{\rm{eff}}=a+a^\prime-1/2>1 \ .
\end{gather}
The latter model in the above range of the parameters corresponds to the following wave-function decay:
\begin{gather}
|\psi_{E_l}(m+R)| \sim
\left\{
  \begin{array}{ll}
    N^{-\gamma/2}e^{-R/N^\gamma}, & R\ll N^{\gamma}\ln N; \\
    N^{-\gamma/2}/R^a, & R\gg N^{\gamma}\ln N.
  \end{array}
\right.
\end{gather}
yielding
\begin{gather}
D_q =
\left\{
  \begin{array}{ll}
    \gamma, & q>\frac{1-\gamma}{2 a_{\rm{eff}}}; \\
    \frac{1-q(2a_{\rm{eff}}+\gamma)}{1-q}, & q<\frac{1-\gamma}{2 a_{\rm{eff}}}.
  \end{array}
\right.
\end{gather}

The main difference between our model and the above one is that the energy level differences are correlated and therefore decay slower than $1/R$. This, in particular, forms the effective dimension $d_{\rm{eff}}=3/2-a^\prime$ for the short-range Anderson model, controlling the limiting $D_2$ at large $a$: cf. $D_{q} = d_{\rm{eff}} \gamma$ with Eq.~\eqref{eq:D2_1d_ALT}.

Moreover, these correlations make the wave-function decay before the perturbative power law $1/R^a$ to be much less clear than the exponential decay in~\cite{Tang2022nonergodic}.

\end{widetext}

\end{document}